\documentclass[aps,prl,reprint,groupedaddress]{revtex4-1}

\usepackage{amsmath}
\usepackage{graphicx}
\usepackage{verbatim}
\usepackage{textcomp}

\begin{document}

\newcommand{\vect}[1]{\mathbf{#1}}

\title{Equivalent Resistance from the Quantum to the Classical Transport Limit}

\author{Saheli Sarkar}
\author{Damaris Kr\"{o}ber}
\author{Dirk K. Morr}
\email[]{dkmorr@uic.edu}

\affiliation{University of Illinois at Chicago, Chicago, IL 60607, USA}

\date{\today}

\begin{abstract}
We generalize the concept of equivalent resistance to the entire range from coherent quantum to diffusive classical transport by introducing the notion of transport equivalent networks. We show that this novel concept presents us with a platform to simplify the structure of quantum networks while preserving their global and local transport properties, even in the presence of electron-phonon or electron-electron interactions. This allows us to describe the evolution of equivalent quantum networks to equivalent classical resistor networks with increasing interaction strength.
\end{abstract}

\maketitle

The equivalent resistance of a network of classical resistors is one of the most fundamental concepts that is used in many fields ranging from physics to engineering. It possesses two characteristic properties: (i) an equivalent resistor leaves the transport properties of the resistor network it replaces unchanged, and (ii) it “simplifies” the structure of the original network \cite{Wu16}. With the continued miniaturization of electronic circuits \cite{Lu01}, the exploration of transport properties at the atomic scale \cite{Web02,Fuec03,Zwan04}, and the ability to design quantum structures at the nanoscale \cite{Gom05,Sing06,Topi07}, it has become of paramount importance to explore whether this concept can be extended not only to the crossover region between classical to quantum transport \cite{Morr08} but indeed to the limit of full quantum coherence. Such an extension could provide a unique venue to extend Moore’s law \cite{Moor09} into the regime of quantum transport, opening unprecedented opportunities for the creation of novel transport functionalities.

In this Letter, we demonstrate that the concept of a classical equivalent resistance can be generalized to the entire range from quantum to classical transport by introducing the concept of transport equivalent networks (TENs). To this end, we describe the
electronic structure of a system in terms of a quantum network \cite{Cres17,Todo18,Todo19} [see Fig.1(a)] and define two networks to be transport equivalent, if they possess identical \textit{IV}-curves for any applied gate voltage. For transport equivalence between networks to exist, it is a sufficient condition that the networks' Hamiltonians are connected by a unitary transformation. This satisfies requirement (i) of an equivalent resistor. To address requirement (ii) -- the simplification of a network -- we note that in a network such as the one shown in Fig.1(a), there exist electronic states that do not take part in charge transport since they possess zero spectral weight at the sites that are connected to the leads. Such states can therefore be eliminated from the network, leading to a simpler structure, without modifying its transport properties. We will show that transport equivalence holds even in the presence of electron-phonon or electron-electron interactions, allowing us tune networks from the non-interacting quantum to the classical transport regime while maintaining their transport equivalence. This enables a mapping of transport equivalent quantum networks onto transport equivalent classical resistor networks, and generalizes the concept of an equivalent resistor to the entire regime from quantum to classical transport.

\begin{figure}[h]
\includegraphics[width=8.cm]{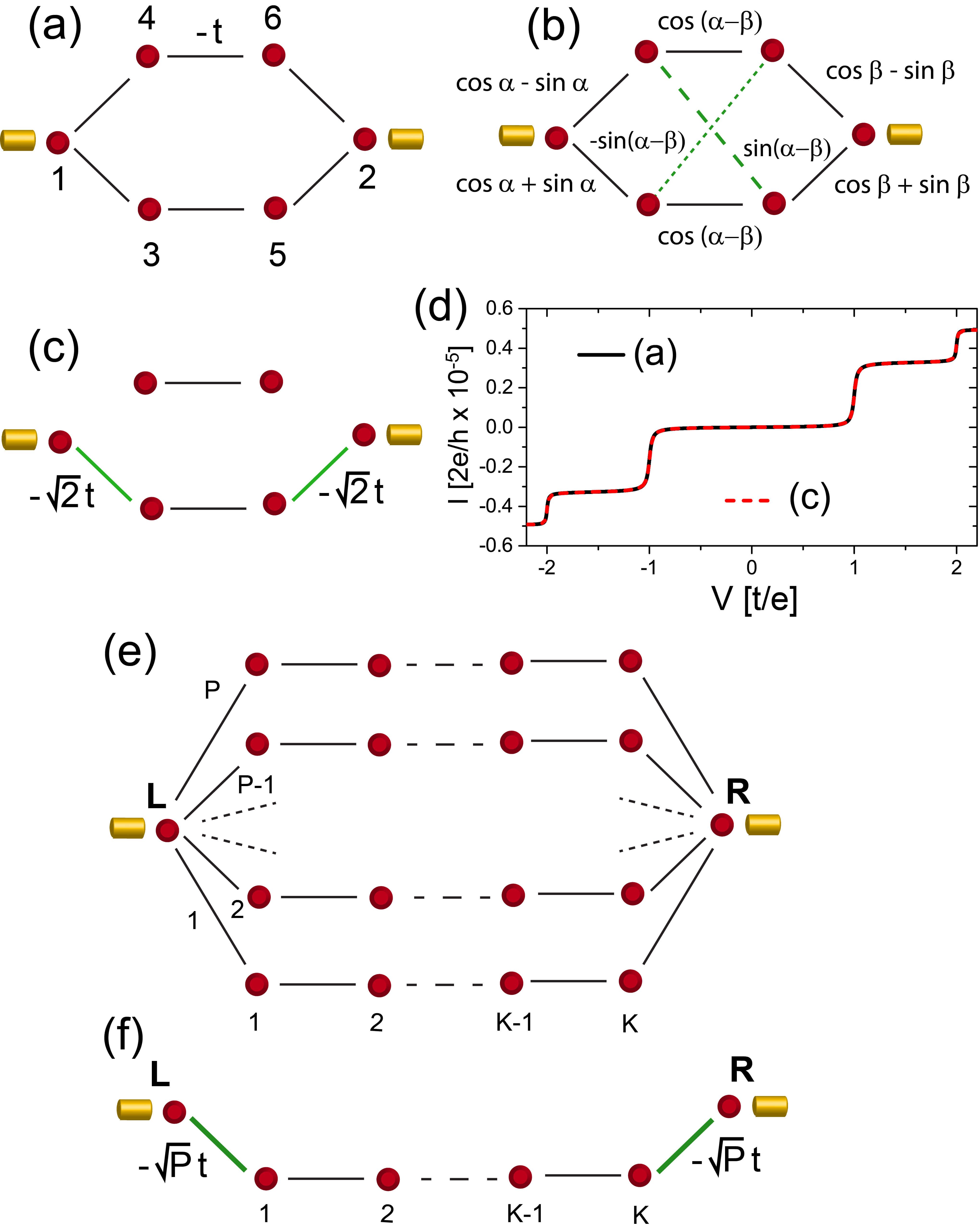}%
\caption{(a) Quantum network with two parallel branches (black lines represent a hopping $-t$). (b) TEN to (a) for arbitrary $\alpha, \beta$ (hopping elements are given in units of $-t$).(c) Simplest transport equivalent network to (a) for $\alpha = \beta = \pi/4$. (d) $IV$-curves for the networks in (a) and (c).(e) Multi-branch network, and (f) its simplest TEN (green lines represent a hopping of $-\sqrt{Pt}$).
\label{Fig 1}}
\end{figure}

To generalize the concept of an equivalent resistor, we represent the electronic structure of a system as a network \cite{Pel15,Cres17,Todo18,Todo19,Figg20} of connected sites that is attached to two (or more) leads, as shown in Fig.\ref{Fig 1}(a), described by the Hamiltonian
$ H = H_{0} + H_{ee} + H_{ph} + H_{lead}$
where
\begin{align}
H_{0} = \sum _{i,j,\sigma}\left( - t_{ij} - E_{0} \delta_{ij}\right)c_{i,\sigma}^{\dagger}c_{j,\sigma} - t_{l} \sum_{{\rm r},i,\sigma}\left(d_{{\rm r},\sigma}^{\dagger}c_{i,\sigma} + h.c.\right) \ .
\end{align}
Here $-t_{ij}$  and $E_{0}$  are the hopping amplitude between sites $i$ and $j$ in the network, and the local on-site energy, respectively, and $c_{i,\sigma}^{\dagger}$ ($d_{i,\sigma}^{\dagger}$) creates a fermion with spin $\sigma$ at site $i$ in the network (leads). The second term describes the coupling of the network to \textit{M} leads with hopping amplitude $-t_{l}$. Moreover, $H_{ee}$ and  $H_{ph}$ describe the electron-electron and electron-phonon interactions in the system (to be discussed below) and $H_{lead}$ represents the electronic structure of the leads. For the subsequent discussion, it is beneficial to rewrite $H_{0}$ in matrix form as $H_{0}=\sum_{\sigma}\Psi_{\sigma}^{\dagger}\hat{H_{0}}\Psi_{\sigma}$ with
$
\Psi_{\sigma}^{\dagger} = \left(d_{1\sigma}^{\dagger},\ldots,d_{\textit{M}\sigma}^{\dagger},c_{1\sigma}^{\dagger},\ldots,c_{\textit{N}\sigma}^{\dagger}\right)
$
being a spinor with the indices of its elements running over all sites in the leads and network, and ${\hat H}_{0}$ being the Hamiltonian matrix.  To compute the charge transport in such networks, we employ the non-equilibrium Keldysh Green's function formalism \cite{Keld21,Caro22,Ram23} where the charge current between sites \textit{i} and \textit{j} in the network is given by
\begin{equation}
I_{ij} = -2\textit{g}_{\textit{s}}\dfrac{\textit{e}}{\hbar}\int_{-\infty}^{\infty} \dfrac{d\omega}{2\pi}\left(-t_{ij}\right)Re\left[G_{ij}^{<}(\omega)\right]  ,
\end{equation}
with $G_{ij}^{<}(\omega)$ being the full lesser Green's function (see supplemental information (SI) Sec.~I), and $\textit{g}_{s} = 2$ representing the spin degeneracy. A current is induced by applying different chemical potentials, $\mu_{L,R} = \pm e \Delta V/2$ in the left (L) and right (R) leads, resulting in a voltage bias $\Delta V$ across the network.

A sufficient condition for two networks to be transport equivalent is that their respective Hamiltonian matrices, $\hat{H_{0}}$ and $\hat{H_{0}^{'}}$ are related by a unitary transformation $\hat{U}$, i.e.,
$\hat{H} = \hat{U}\hat{H_{0}}\hat{U^{\dagger }}$ (see SI Sec.~II). This equivalence holds irrespective of the specific form of $H_{lead}$. Additionally, we require that the electronic and spatial coupling to the leads be the same for both networks, implying that the $M$ sites, through which the current enters or exits the networks as well as the $M$ lead sites, are unaffected by the unitary transformation. As a result, $\hat{U}$ possesses the following matrix representation
\begin{equation}
\hat{U} =
\left(\begin{array}{c c}
\hat{1}&\hat{0}\\
\hat{0}&\hat{Q}
\end{array}\right)
\end{equation}
where $\hat{1}$ is the $2M \times 2M$  identity matrix that acts on the $M$ network sites and the $M$ leads that are connected, and $\hat{Q}$ is an $(N-M) \times (N-M)$  unitary matrix that acts on all other sites of the network.

To exemplify the concept of transport equivalent networks, we consider a network with two parallel branches [Fig.~\ref{Fig 1}(a)] which represents the quantum analog of a classical parallel resistor network. Using the numbering of sites shown in Fig.~\ref{Fig 1}(a), the unitary transformation
\begin{equation}
\hat{U} =
\left(\begin{array}{c c c}
\hat{1}&0&0\\
0&\hat{D}(\alpha)&0\\
0&0&\hat{D}(\beta)
\end{array}\right)
\end{equation}
where
$
\hat{D}(\alpha) =
\left(\begin{array}{c c}
\cos{\alpha}&-\sin{\alpha}\\
\sin{\alpha}&\cos{\alpha}
\end{array}\right) \
$
yields a whole class of TENs whose electronic hopping elements are parameterized by the angles $\alpha$ and $ \beta$ [see Fig.~\ref{Fig 1}(b)].
While all of these TENs possess identical transport properties, they are in general, however, not “simpler” than the original network since even new hopping elements can emerge. The main challenge therefore lies in finding a transformation $\hat{U}$ that yields the largest possible simplification of a network, with the exact meaning of simplification being dependent on the particular network properties one might be interested in. For example, the network of Fig.~\ref{Fig 1}(b) can be simplified to a smaller network by choosing $\alpha=\beta=\pi/4$  [Fig.~\ref{Fig 1}(c)]. In this case the TEN separates into two disjoint parts, of which only one is connected to the leads and thus contributes to charge transport.   The disconnected part can therefore be omitted, implying that the original network [Fig.~\ref{Fig 1}(a)] consisting of six sites can be replaced by a simpler (i.e., smaller) network with identical transport properties that consists of four sites with renormalized hopping amplitudes [Fig.~\ref{Fig 1}(c)]. The \textit{IV} curves of the original network [Fig.~\ref{Fig 1}(a)] and of the TEN [Fig.~\ref{Fig 1}(c)] computed from Eq.(2) are as expected identical [see Fig.~\ref{Fig 1}(d)]. This result can immediately be generalized to a network with $P$ parallel branches each consisting of $K$ sites [Fig.~\ref{Fig 1}(e)] with the smallest TEN possessing only a single branch [Fig.~\ref{Fig 1}(f)] (the explicit form of $\hat{U}$ is given in SI Sec.~III.A). Identifying the unitary transformation that yields the simplest TEN thus represents the quantum analog of finding the classical equivalent resistor.

\begin{figure}[h]
\includegraphics[width=8.cm]{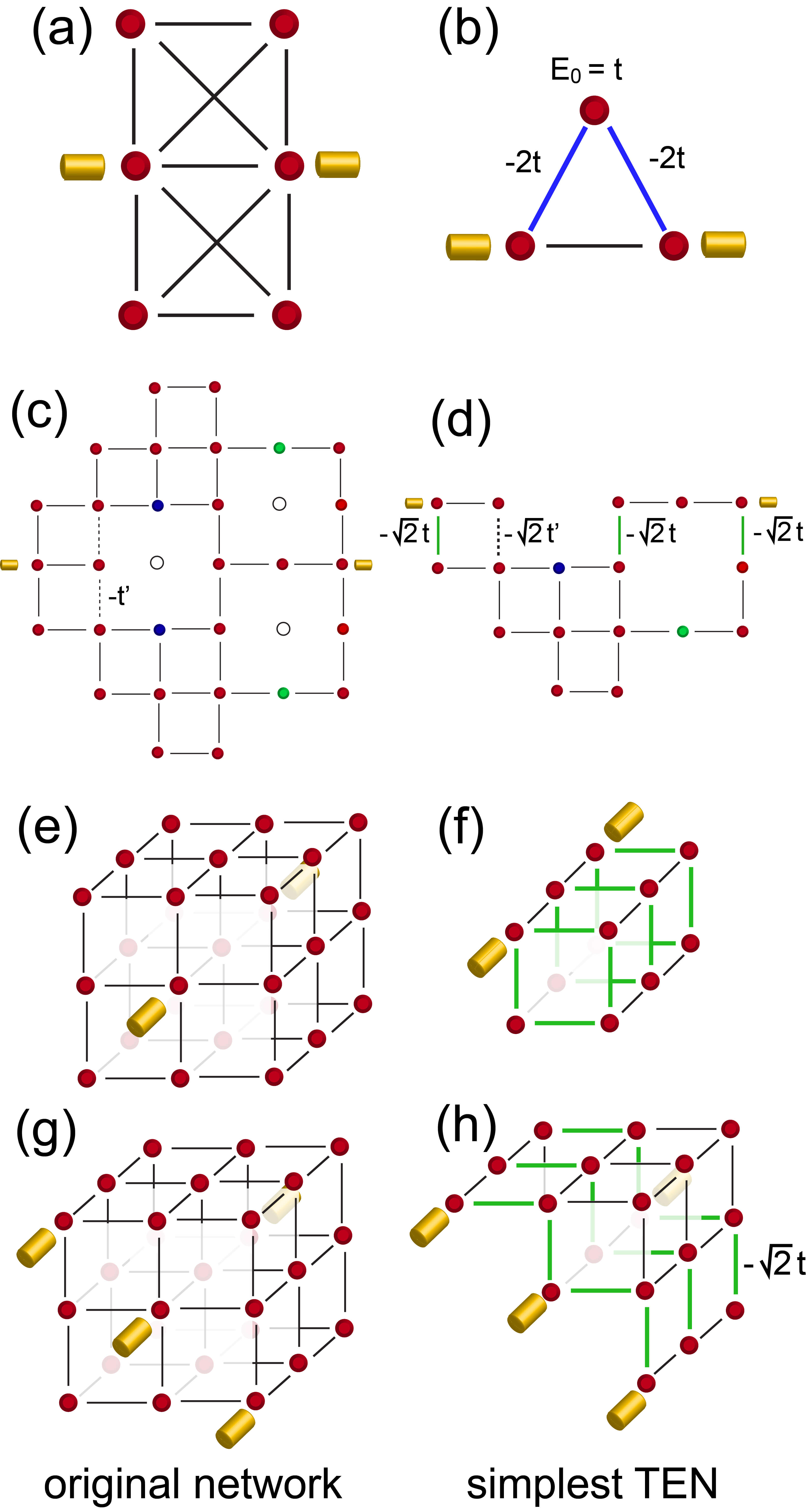}
\caption{(a) Network with next-nearest neighbor hopping, and (b) its simplest TEN (unless otherwise noted, $E_{0} = 0$). (c) Disordered network (open circles represent missing sites, circles of the same color possess the same $E_{0}$, dotted line represents a hopping of $-t'$), and (d) its simplest TEN. Three-dimensional networks with different leads possessing (e) a $C_{4}$, and (g) a $C_{2}$ symmetry around their center axis. (f),(h) Simplest TENs to (e) and (g). Blue and green lines represent hoppings of $-2t$ and $-\sqrt{2}t$, respectively.}
\label{Fig 2}
\end{figure}
The concept of transport equivalence can be applied to a wide variety of networks such as networks containing next nearest-neighbor hopping amplitudes [Figs.~\ref{Fig 2}(a), (b), the corresponding $\hat{U}$ is given in SI Sec.~III.B], networks with a square lattice or graphene lattice structure (see SI Sec.~III.C), networks that include disorder in the on-site energies or hopping amplitudes [Figs.~\ref{Fig 2}(c), (d), see SI Sec.~III.D], or three-dimensional networks [Figs.~\ref{Fig 2}(e) - (h), SI Sec.~III.E].  All of the TENs shown in the right column of Fig.~\ref{Fig 2} are TENs with the smallest number of network sites and hopping elements, thus realizing the greatest simplification of the original networks (left column). The number of networks sites in the smallest possible TEN can be deduced from the original network: if the original network of $N$  sites possesses $N_{0}$  states whose wave-functions simultaneously vanish at all sites that the leads are connected to, then this network can be transformed into a class of “smallest” TENs in which only $S_{min} = N - N_{0}$ sites are connected to the leads, while the remaining $N_{0}$ sites are disconnected and therefore irrelevant for charge transport. For example, the network shown in Fig.~\ref{Fig 1}(e) possesses $(P\times K+2)$ sites and $(P-1)\times K$  states whose wave-function vanishes at sites \textbf{L} and \textbf{R} which are connected to the leads. It can therefore be transformed into a new network with $S_{min} = K + 2$ sites [Fig.~\ref{Fig 1}(f)] that does not only contain the smallest of possible number of sites, but also of hopping elements, and therefore represents the simplest possible TEN. We note that systems (i.e., network and leads) that have a higher symmetry [such as that in Fig.~\ref{Fig 2}(e) with $C_{4}$-symmetry] in general possess TENs with a smaller $S_{min}$ than systems with a lower symmetry [such as that in Fig.~\ref{Fig 2}(g) with $C_{2}$-symmetry].

\begin{figure}[h]
\includegraphics[width=8.cm]{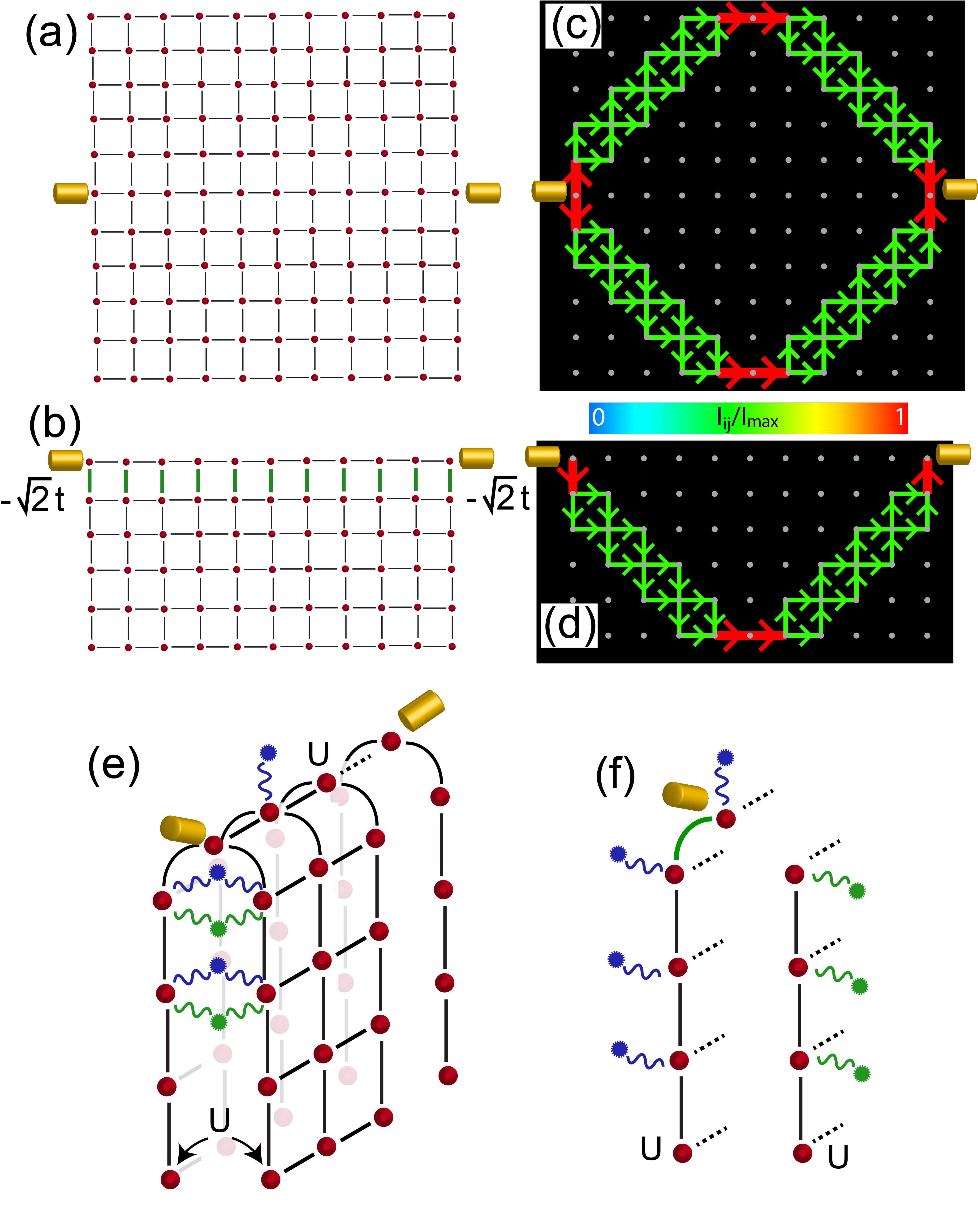}%
\caption{(a) Square-lattice network, and (b) its simplest TEN. (c),(d) Spatial pattern of the normalized currents carried by the $E=0$ state in the networks of (a) and (b) for $\Delta V = 0.01t/e$. (e) "Bent" square lattice network with sites symmetric around the center row interacting with the same two phonon modes (blue and green wavy line) and via an electron-electron interaction $U$. (f) Simplest TEN to (e).}
\label{Fig 3}
\end{figure}
The transport equivalence of networks is also reflected in a close similarity of their spatial current patterns. To demonstrate this, we consider the $N_{x}\times N_{y}$  square lattice network shown in Fig.~\ref{Fig 3}(a) (with $N_{x} = N_{y} = 11$), and its smallest TEN consisting of $N_{x}(N_{y} +1)/2$ sites [Fig.~\ref{Fig 3}(b), see SI Sec.~III.C].
In this TEN, only the vertical hopping elements directly connected to the top row are modified to $-\sqrt{2t}$ in comparison to the original network. The $IV$ curves of these two networks are identical (see SI Sec.~IV), and the spatial patterns of current flow through both networks [see Figs.~\ref{Fig 3}(c) and (d)] exhibit a close similarity, in that the current pattern in the TEN is all but identical to that in the lower half of the original network. This similarity also holds for the current patterns carried by other energy states and in networks with $N_{x} \not = N_{y}$  (see SI Sec.~IV), and thus establishes the equivalence of not only global but also local transport properties.

The concept of transport equivalent networks can also be extended to interacting networks containing electron-electron or electron-phonon interactions. In this case, a simpler TEN exists if the form of the interactions in the original network is such that after applying the unitary transformation, the interaction does not couple the disjoint parts in the TEN. To demonstrate this, we consider a “bent” square-lattice network [Fig.~\ref{Fig 3}(e)] in which electrons on sites $i$ and $j$  (which are symmetric with respect to the center row) interact with the same two phonon modes as described by
\begin{align}
H_{ph} &= \dfrac{g}{2}\sum_{\langle i,j \rangle,\sigma} \left[ \left( n_{i,\sigma} +  n_{j,\sigma}\right) \left(a_{i}^{\dagger} + a_{i} + a_{j}^{\dagger} + a_{j}\right) \right. \nonumber \\
  & \left. + \left( c_{i,\sigma}^{\dagger}c_{j,\sigma} + c_{j,\sigma}^{\dagger}c_{i,\sigma}\right) \left( a_{i}^{\dagger} + a_{i}\ -a_{j}^{\dagger} - a_{j} \right) \right] \nonumber \\
  & +  g \sum_{r,\sigma}n_{r,\sigma}\left(a_{r}^{\dagger}+ a_{r}\right)
 + \sum_{k}\omega_{k}a_{k}^{\dagger}a_{k} \ .
\label{eq:Hph1}
\end{align}
Here, $n_{i,\sigma}=c_{i,\sigma}^{\dagger}c_{i,\sigma}$  is the local occupation operator, $g$ is the interaction strength, and the first sum runs over all pairs of symmetric pairs of sites $i$ and $j$, the second sum runs over the sites of the middle row, and the last one over all phonon modes with energy $\omega_{k}$. In the TEN [Fig.~\ref{Fig 3}(f), using the unitary transformation of Eq.(S32) in SI Sec.~IIIC], the electron-phonon interaction is entirely local and given by
\begin{equation}
H_{ph}^{'} = g \sum_{i,\sigma}\left(a_{i}^{\dagger} + a_{i}\right) + \sum_{k}\omega_{k}a_{k}^{\dagger}a_{k} \ .
\label{eq:Hph2}
\end{equation}
Thus, the unitary transformation yields two disjoint parts of the TEN, even in the presence of the electron-phonon interactions and we can again neglect the part disconnected from the leads. Similarly, an electron-electron interaction in the original network of the form [Fig.~\ref{Fig 3}(e)]
\begin{align}
H_{ee} &= \dfrac{U}{2}\sum_{\langle i,j \rangle}(c_{i \uparrow}^{\dagger}c_{j \uparrow}c_{i \downarrow}^{\dagger}c_{j \downarrow} + c_{i \uparrow}^{\dagger}c_{j \uparrow}c_{j \downarrow}^{\dagger}c_{i \downarrow}) \nonumber \\
& + \dfrac{U}{2} \sum_{\alpha,\beta =\langle i,j \rangle} n_{\alpha\uparrow} n_{\beta\downarrow}+ U \sum_{r} n_{r \uparrow}n_{r \downarrow}
\end{align}
transforms into a purely local Coulomb interaction in the TEN [Fig.~\ref{Fig 3}(f)]
\begin{equation}
H_{el} = U \sum_{i} n_{i \uparrow} n_{i \downarrow}
\end{equation}
allowing us to again neglect the disjoint part of the TEN.

\begin{figure}[h]
\includegraphics[width=8.cm]{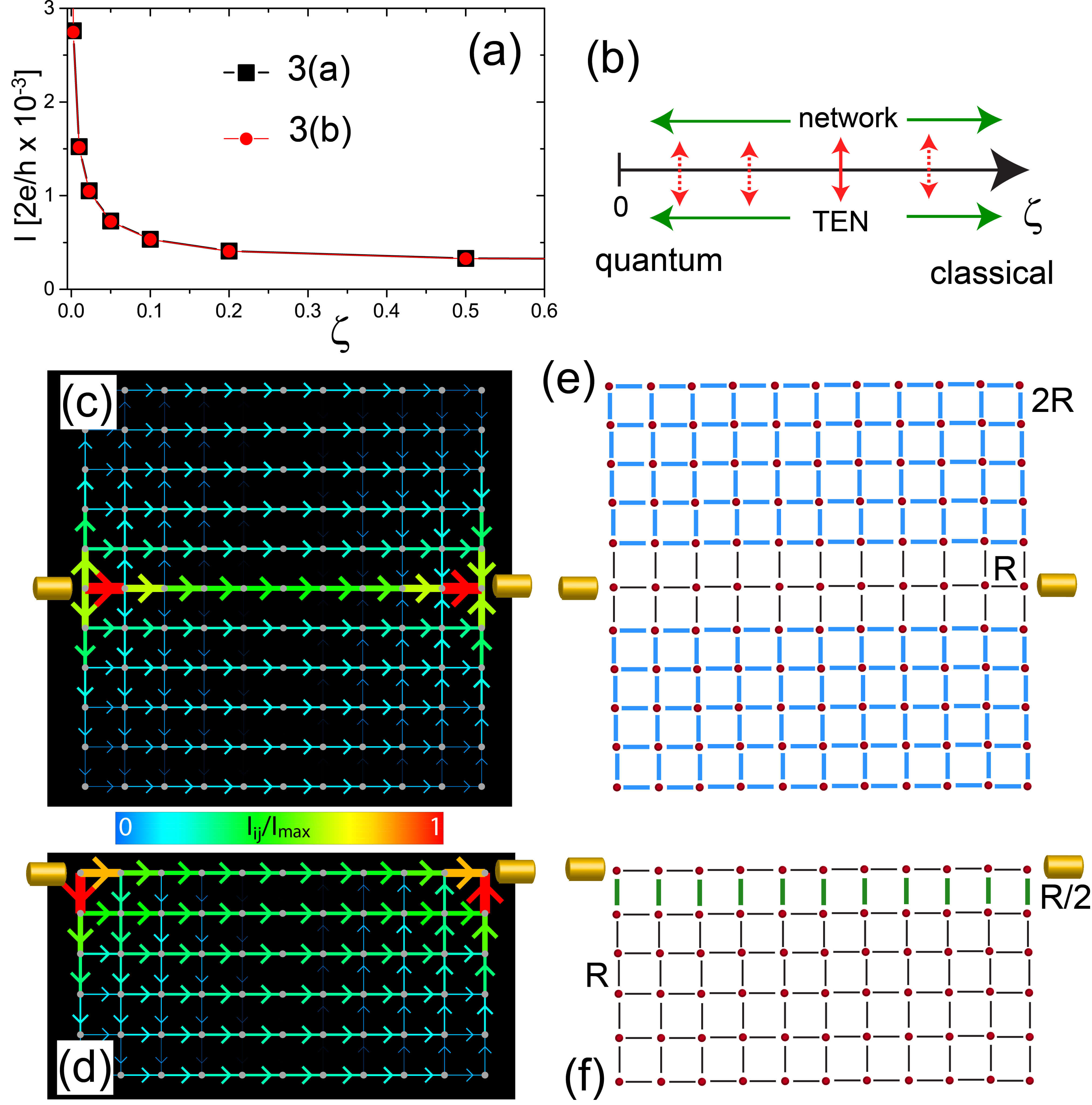}
\caption{(a) Current through the networks of Fig.~\ref{Fig 3}(a) and (b) as a function of $\zeta$ for $\mu_{L,R} = \pm 0.005t$. (b) Schematic representation of the transport equivalence between networks holding over the entire range of $\zeta$ from the quantum to the classical transport regime. (c),(d) Spatial pattern of the normalized current carried by the $E = 0$ state in the networks of Figs.~\ref{Fig 3}(a) and (b) for $\Delta V = 0.0004t/e$ and $\zeta = 50 t^{2}$, approaching the classical limit $\zeta\rightarrow \infty$. (e),(f) Classical resistor networks that possess the same spatial current patterns as in (c) and (d). Black, blue and green lines represent a resistance of $R$, $2R$, and $R/2$, respectively.
\label{Fig 4}}
\end{figure}
To demonstrate the transport equivalence of networks in the presence of an electron-phonon interaction, we consider the Hamiltonians of Eqs.(\ref{eq:Hph1}) and (\ref{eq:Hph2}) in the networks of Figs.~\ref{Fig 3}(a) and (b), respectively. The computational evaluation of their transport properties is in general quite demanding for arbitrary temperature and phonon energy $\omega_{k}$. We therefore simplify this task by considering a single phonon energy $\omega_{k} = \omega_{0}$ in the high-temperature limit  $k_{B} T \gg \hbar \omega_{0}$ \cite{Bih24} (see SI Sec.~I) yielding an effective electron-phonon interaction given by $\zeta = 2 g^{2} k_{B} T /\left(\hbar \omega_{0}\right)$.  Not only is the resulting total current through these networks identical for all $\zeta$ [Fig.~\ref{Fig 4}(a)], but the close similarity of the spatial current patterns in the TENs also persists over the entire range from the quantum $\zeta = 0$  [Figs.~\ref{Fig 3}(c) and (d)] to the classical $\zeta \rightarrow \infty$  transport limit \cite{Morr08} [Figs.~\ref{Fig 4}(c) and (d), SI Sec.~V]. In the latter limit, the TENs' current patterns are identical to those of the (equivalent) classical resistor networks shown in Figs.~\ref{Fig 4}(e) and (f) (SI Sec.~VI). Thus, the networks' transport equivalence holds for all strengths of the electron-phonon interaction, as schematically shown in Fig.~\ref{Fig 4}(b), demonstrating that the concept of transport equivalent networks can be extended to the entire range from quantum to classical transport.

 We note that the above generalization of the equivalent resistance should be applicable to any system in which transport can be described within the above formalism, such as networks of quantum dots \cite{Tala10,Parth11}, molecules and polymers \cite{Moul12}, and excitonic energy transfer networks in light–harvesting complexes \cite{Sch13,Fle14,Pel15}. This clearly establishes the wide-ranging importance and appeal of this novel concept in exploring and discovering transport phenomena in a broad range of materials and systems. Finally, an interesting extension of the above work is to relax the requirement of exact transport equivalence, and to consider networks that are ``nearly" transport equivalent. We expect that this would allow us to further reduce the size of networks, and to consider simpler forms of interactions.

\begin{acknowledgments}
We would like to thank T. Can, D. Goldhaber-Gordon, and M. Vojta for helpful discussions. This project was supported by  the U. S. Department of Energy, Office of Science, Basic Energy Sciences, under Award No. DE-FG02-05ER46225.
\end{acknowledgments}

\end{document}


\fontsize{11}{13}

\begin{center}
{\large {\bf  Supplemental Online Information for}} \\[0.5cm]
\end{center}

\title{Equivalent Resistance from the Quantum to the Classical Transport Limit}

\author{Saheli Sarkar, Damaris Kr\"{o}ber, and  Dirk K. Morr}

\affiliation{Department of Physics, University of Illinois at Chicago, Chicago,
IL 60607}

\date{\today}

\maketitle

\section{Charge Transport through Networks and the Non-Equilibrium Keldysh Formalism}
\label{sec:method}

To investigate the charge transport through networks, we consider a network of $N$ sites that is coupled to $M$ leads. We assume that each lead is coupled to a single site in the network (and vice versa) such that there are $M$ sites in the network that are coupled to the $M$ leads. The entire systems is described by the Hamiltonian $H=H_0 + H_{ph} +H_{lead}$ where
\begin{align}
H_0=&\sum_{i,j,\sigma} \left( -t_{ ij} - E_0 \delta_{ ij} \right) c^\dagger_{ i,\sigma} c_{ j,\sigma} -t_l \sum_{r,i,\sigma} \left ( d^\dagger_{ r,\sigma} c_{ i,\sigma} + H.c. \right) = \sum_{\sigma} \Psi^\dagger_{\sigma} {\hat H}_0 \Psi_{\sigma} \ .
\label{eq:H0}
\end{align}
Here, $c^\dagger_{ i,\sigma} (d^\dagger_{ r,\sigma})$ creates an electron with spin $\sigma$ at site $i$ in the network (site $r$ in the leads), $-t_{ij}$ is the electronic hopping between sites $i$ and $j$ in the network, $E_0$ is the on-site energy (unless otherwise noted, we set in the following $E_0=0$), and $-t_l$ is the hopping amplitude between the network and the leads. It is beneficial to write $H_0$ in matrix form by introducing the spinors
\begin{equation}
\Psi_\sigma^{\dagger }=\left(
\begin{array}{c}
d_{1,\sigma}^{\dagger } \\
\vdots  \\
d_{M,\sigma}^{\dagger } \\
c_{1,\sigma}^{\dagger } \\
\vdots  \\
c_{M,\sigma}^{\dagger } \\
c_{M+1,\sigma}^{\dagger } \\
\\
c_{N,\sigma}^{\dagger }%
\end{array}%
\right) \ , \qquad \Psi =\left(
\begin{array}{ccccccccc}
d_{1,\sigma}, & \ldots , & d_{M,\sigma}, & c_{1,\sigma}, & \ldots , & c_{M,\sigma}, &
c_{M+1,\sigma}, & \ldots , & c_{N,\sigma}
\end{array}%
\right)
\label{eq:spinor}
\end{equation}
and ${\hat H}_0$ is the corresponding Hamiltonian matrix in Eq.(\ref{eq:H0}). The first $M$ entries in the spinor are the lead operators, and the
second $M$ entries are the operators for the sites that are connected to the leads. The electronic structure of the leads is described by $H_{leads}$, which, however, is irrelevant for the current discussion. Moreover, we consider a generalized electron-phonon interactions of the form
\begin{align}
H_{ph} &=  \sum_{i,j,k, \sigma} g_{ij}^{(k)} c^\dagger_{i,\sigma} c_{j,\sigma}  \left( a_k^\dagger + a_k \right) + \sum_{k} \omega_k  a^\dagger_k a_k
\label{eq:phonon}
\end{align}
Here, $a^\dagger_k$ creates a phonon in the $k'th$ mode with phonon energy $\omega_k$. Moreover, $g^{(k)}_{i,j}$ is the electron-phonon coupling strength, leading to the scattering of an electron between sites $j$ and $i$ of the network by the $k'th$ phonon mode. Finally, disorder in the network can arise either from disorder in the hopping elements, $-t_{ij}$, or from disorder due to scattering of the network's electrons by non-magnetic defects as described by the Hamiltonian
\begin{equation}
H_U =  {\sum_{i,\alpha}} U_{i} c^\dagger_{i,\alpha} c_{i,\alpha}
\label{eq:U0}
\end{equation}
where $U_{i}$ is the non-magnetic scattering potential at site $i$, and the sum runs over all defect locations (we here assume point-like scatterers). This scattering potential is equivalent to spatial variations in the on-site energy $E_0$, and therefore can be included in the definition of ${\hat H}_0$ in Eq.(\ref{eq:H0}).

To investigate the flow of charge through the network, we employ the non-equilibrium Keldysh Green's function formalism \cite{Kel65,Car71a}. Within this formalism, the current between sites $i$ and $j$ in the network is induced by different
chemical potentials, $\mu_{L,R}=\pm V_0/2$ in the left and right leads, and given by \cite{Car71a}
\begin{equation}
I_{ij}=-2 \frac{e}{\hbar}  g_s \;
\intop_{-\infty}^{+\infty}\frac{d\omega}{2\pi} \left( -t_{ij} \right) {\rm Re} \left[
G^<_{ij}(\sigma, \omega)\right] \ . \label{eq:Current}
\end{equation}
with $g_s=2$ representing the spin degeneracy of the network, and $G^<_{ij}(\sigma, \omega)$ being the full, non-local lesser Green's function, defined via ${\hat G}^<_{ ij}(\sigma,t,t) = i \langle c^\dagger_{i,\sigma}(t)  c_{j,\sigma}(t) \rangle $ in the time domain. To account for the effects of  electronic
hopping, the presence of defects, the electron-phonon interaction, and the coupling to the leads, we employ the Dyson equations for the lesser and retarded Green's functions. By defining lesser and retarded Green's function matrices $\hat{G}^{<,r}$ in real space whose $(ij)$-elements are given by $\hat{G}^{<,r}_{\ij}$, we obtain the Dyson equations in frequency space
\begin{subequations}
\begin{align}
\hat{G}^{<} &= \hat{G}^{r}\left[ \left(\hat{g}^{r} \right)^{-1} \hat{g}^{<}
\left( \hat{g}^{a} \right)^{-1} + {\hat \Sigma}^<_{ph} \right] \hat{G}^{a} \label{eq:fullGa} \\
\hat{G}^{r} &= \hat{g}^{r} + \hat{g}^{r} \left[ \hat{H}_0 + {\hat \Sigma}^r_{ph}
\right] \hat{G}^{r} \ .
\label{eq:fullGb}
\end{align}
\end{subequations}
Here, ${\hat \Sigma}^{r,<}_{ph}$ are the retarded and lesser fermionic self-energy matrices arising from the
electron-phonon interaction, and $\hat{g}^{r,a,<}$ are the retarded, advanced and lesser fermionic Green's function matrices of the network and the leads in the absence of any electronic hopping, defect scattering or electron-phonon interaction. These Green's functions are given by $(x=r,a,<)$
\begin{equation}
\hat{g}^{x}= \left(
\begin{array}{cc}
\hat{g}_{leads}^{x} & 0  \\
0 & \hat{g}^{x}
\end{array}%
\right)
\end{equation}
where $\hat{g}^{x}$ and $\hat{g}_{leads}^{x}$ are the Green's function matrices describing the network and the leads, respectively. Moreover, $\hat{g}^{x}$ are diagonal matrices with elements
\begin{subequations}
\begin{align}
g_0^r(\omega) &= \frac{1}{\omega + i \delta - eV_g} \\
g_0^<(\omega) &= -2i n_F(\omega) {\rm Im} g_0^r(\omega)
\end{align}
\end{subequations}
where $n_F(\omega)$ is the Fermi distribution function, $e$ is the electron charge and $V_g$ is the gate voltage. Note that to move a state from energy $E_i>0$ to the Fermi energy, one has to apply the gate voltage $V_g=E_i/e$. Moreover, $\hat{g}_{leads}^{x}$ are diagonal matrices with elements
\begin{subequations}
\begin{align}
g_{leads}^r(\omega) &= -i \pi \\
g_{leads}^<(\omega) &= -2i \  n_F(\omega+\mu_{L,R}) \ {\rm Im} g_0^r(\omega)
\end{align}
\end{subequations}
implying that the leads' density of states is equal to unity and that we consider the wide band limit for the leads. Moreover, $\mu_{L,R}$ is the chemical potential in the left and right leads, giving rise to a potential difference $\Delta V=(\mu_L - \mu_R)/e$ across the network. The spin-resolved local density of states, $N_\sigma(i, E)$ at site $i$ and energy $E$ is obtained from Eq.(\ref{eq:fullGb}) via
\begin{equation}
N_\sigma(i, E=\hbar \omega) = -\frac{1}{\pi} {\rm Im} \ {\hat G}_{ii}(\omega) \ .
\end{equation}

To study the effects of dephasing on transport equivalent networks, we consider the electron-phonon interaction of Eq.(\ref{eq:phonon}).
The computation of the fermionic self-energy ${\hat \Sigma}_{ph}$ arising from such an electron-phonon interaction within a conserving-approximation is computationally quite demanding for extended networks and arbitrary temperature $T$ and phonon energy $\omega_k$, but can be significantly simplified in the high-temperature limit $k_B T \gg \hbar \omega_k$ $\forall k$ where one can make use of the high-temperature approximation
introduced in Ref. ~\cite{Bih05}).  In this case, one retains only those terms in ${\hat \Sigma}_{ph}$ that contain the Bose distribution function since in this limit $n_B(\omega_k)\gg 1$.  The fermionic self-energy can best be written in matrix form, whose $(ij)$ elements are given by $\Sigma_{ij}^{r,<}(\omega)$. Within the self-consistent Born approximation, the self-energy is computed using the full Green's function of Eqs.(\ref{eq:fullGa}) and (\ref{eq:fullGb}), given by
\begin{align}
\Sigma_{ij}^{r,<}(\omega) =  i \sum_{r,s,k} g^k_{ir} g^k_{s,j} \int \frac{d\nu}{2\pi} D_k^{<}(\nu)
G_{rs}^{r,<}( \omega - \nu) \ ,
\label{eq:SE}
\end{align}
where
\begin{subequations}
\begin{align}
D_k^{<}(\omega) =& 2 i n_{B}(\omega) {\rm Im} D_k^{r} (\omega)  \\
D_k^{r}(\omega) =& \frac{1}{\omega - \omega_k + i\delta} - \frac{1}{\omega + \omega_k + i\delta}
\end{align}
\end{subequations}
are the lesser and retarded phonon Green's functions, which we assume to remain unchanged in
the presence of an applied bias. In the following, we assume for simplicity that the energy $\omega_k$ of all phonon modes $k$ is the same, i.e., $\omega_k=\omega_0$ $\forall k$, and that all interaction energies, $g_{rj}=g$ are the same. Finally, we consider the limit $\omega_{0} \rightarrow 0$ in which the self-energy, to leading order in
$k_{B}T/\hbar \omega_{0}$, becomes
\begin{align}
\Sigma_{ij}^{r,<}(\omega) &=  2 \sum_{r,s,k} \left( \frac{g^k_{ir}}{g} \right) \left( \frac{g^k_{sj}}{g} \right)  g^{2} \frac{k_{B}T}{\hbar \omega_0} G_{rs}^{r,<}(\omega) \equiv \sum_{r,s,k} \left( \frac{g^k_{ir}}{g} \right) \left( \frac{g^k_{sj}}{g} \right)  \zeta G_{rs}^{r,<}(\omega) \ .
\label{eq:sigma}
\end{align}
Using the self-energy given in Eq.(\ref{eq:sigma}), we then first self-consistently compute the retarded Green's function matrix given in Eq.(\ref{eq:fullGb}), and subsequently the lesser Green's function matrix in Eq.(\ref{eq:fullGa}). Note that as
\begin{align}
\frac{g^k_{ir}}{g} = \left\{
\begin{array}{l}
1  \qquad \text{if
}  \ \ g^k_{ir} \not = 0 \\
0 \qquad \text{otherwise }%
\end{array}%
\right. \qquad ,
\end{align}
the effective strength of the electron-phonon interaction is given by $\zeta = 2 g^{2} \frac{k_{B}T}{\hbar \omega_0}$.

\section{Invariance of Transport Properties under Unitary Transformations}

In this section, we demonstrate analytically that the transport properties of the network (i.e., the total current flowing through a network for a given applied bias and gate voltage) remain unchanged under a unitary transformation of the form
\begin{equation}
\hat{U}= \left(
\begin{array}{cc}
{\hat 1} & 0  \\
0 & \hat{Q}
\end{array}
\right)
\label{eq:U}
\end{equation}
where ${\hat 1}$ is the $(2M \times 2M)$ identity matrix that acts on the $M$ network and $M$ lead sites that are connected to each other, and ${\hat Q}$̂ is an $(N-M) \times (N-M)$ unitary matrix that acts on all other sites of the network. To this end, we have to show that the current flowing between any of the lead sites and the network  remains unchanged under the unitary transformation. To this end, we denote an arbitrary lead site by $\alpha$, and the network site it is connected to by $\beta$, such that the current flowing between these two sites is given by
\begin{equation}
I_{\alpha \beta}=-2 \frac{e}{\hbar} g_s \;
\intop_{-\infty}^{+\infty}\frac{d\omega}{2\pi} (-t_l) {\rm Re} \left[
G^<_{\alpha \beta }(\sigma, \omega)\right] \ . \label{eq:Current}
\end{equation}
It is therefore sufficient to show that $G^<_{\alpha \beta}(\sigma, \omega)$ remains invariant under the unitary transformation.

We next define the lesser Greens function matrix in real time via
\begin{equation}
\hat{G}^{<}(\sigma,t,t)=i\left\langle \Psi_\sigma^{\dagger }(t)\Psi_\sigma (t)\right\rangle
\end{equation}
with spinors $\Psi_\sigma^{\dagger }, \Psi_\sigma$ defined in Eq.(\ref{eq:spinor}). Under the unitary transformation of Eq.(\ref{eq:U}), the spinors transform as
\begin{subequations}
\begin{align}
\Phi_\sigma^{\dagger }(t) &=\hat{U}^{\dagger }\Psi_\sigma^{\dagger }(t) \\
\Phi_\sigma (t) &=\Psi_\sigma (t)\hat{U}
\end{align}
\end{subequations}
where $\Phi_\sigma^{\dagger }(t)$ is a spinor containing the new operators in the TEN. Thus one obtains for
the Green's function matrix in the TEN
\begin{eqnarray}
\hat{G}^{\prime <}(\sigma,t,t) &=&i\left\langle \Phi_\sigma^{\dagger }(t)\Phi_\sigma
(t)\right\rangle =i\left\langle \hat{U}^{\dagger }\Psi_\sigma^{\dagger }(t)\Psi_\sigma (t)%
\hat{U}\right\rangle =\hat{U}^{\dagger }i\left\langle \Psi_\sigma^{\dagger
}(t)\Psi_\sigma (t)\right\rangle \hat{U} \nonumber \\
&=&\hat{U}^{\dagger }\hat{G}^{<}(\sigma,t,t)\hat{U}
\end{eqnarray}%
Let us now consider an element of $\left[ \hat{G}^{\prime <}(\sigma,t,t)\right]
_{\alpha \beta }$ between the lead site $\alpha $ and a site $\beta $ coupled
to that lead%
\begin{equation}
\left[ \hat{G}^{\prime <}(\sigma,t,t)\right] _{\alpha \beta }=\left[ \hat{U}%
^{\dagger }\hat{G}^{<}(\sigma,t,t)\hat{U}\right] _{\alpha \beta }=\sum_{\gamma
\delta }\hat{U}_{\alpha \gamma }^{\dagger }\hat{G}_{\gamma \delta }^{<}(\sigma,t,t)%
\hat{U}_{\delta \beta }
\end{equation}
Since the operators at sites $\alpha $ and $\beta $ are remain unchanged
under the unitary transformation, we have
\begin{align}
\hat{U}_{\alpha \gamma }^{\dagger } =&\delta _{\alpha \gamma } \nonumber \\
\hat{U}_{\delta \beta } =&\delta _{\delta \beta }
\end{align}%
which yields%
\begin{equation}
\left[ \hat{G}^{\prime <}(\sigma,t,t)\right] _{\alpha \beta }=\left[ \hat{G}%
^{<}(\sigma,t,t)\right] _{\alpha \beta } .
\end{equation}
Thus $\hat{G}^{<}_{\alpha \beta}$ is invariant under the above unitary transformation, implying that the total current through the system does not change. As this holds for arbitrary chemical potentials in the leads, and arbitrary gate voltage, we conclude that two networks whose Hamiltonians are connected by a unitary transformation of the form shown in Eq.(\ref{eq:U}), are transport equivalent.

\section{Examples of Unitary Transformations}

In this section, we present the explicit form of the unitary transformations that we used to derive the transport equivalent networks discussed in the main text.

\subsection{Quantum Network with Parallel Branches}

\begin{figure}[h]
 \begin{center}
\includegraphics[width=14.cm]{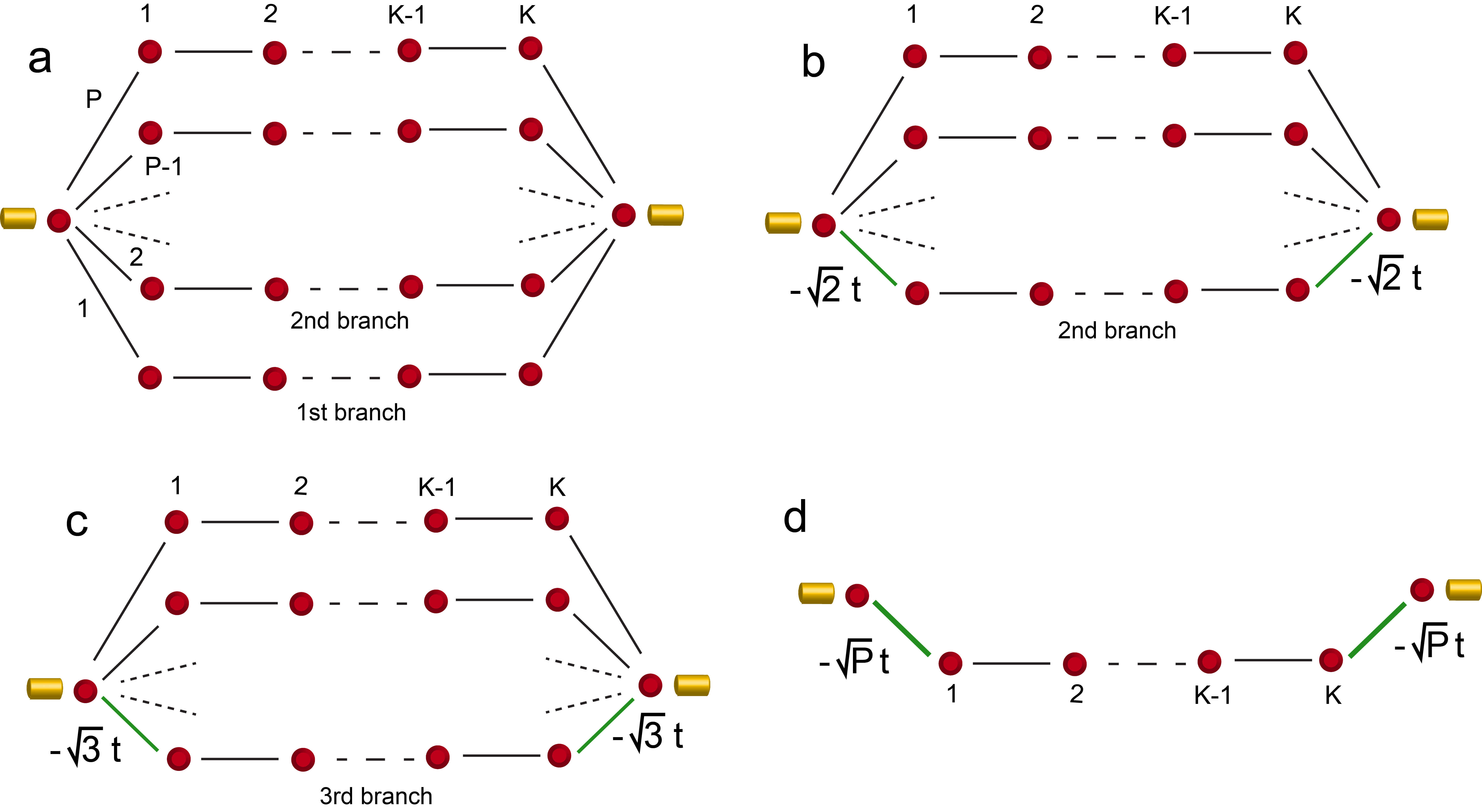}
\caption{(a) Network consisting of $P$ parallel branches each possessing $K$ sites. (b) Transport equivalent network resulting from applying the unitary transformation ${\hat U}_{1,2}$ to the network in (a). (c) Transport equivalent network resulting from applying the unitary transformation ${\hat U}_{1,2} \cdot {\hat U}_{2,3}$ to the network in (a). (d) Simplest transport equivalent network resulting from applying the unitary transformation ${\hat U}_{tot}$ to the network in (a). }
\label{fig:unitary_trans_1}
\end{center}
\end{figure}

We begin by considering a quantum network consisting of $P$ branches with $K$ sites each, as shown in Fig.~\ref{fig:unitary_trans_1}(a) (see also Fig.1{\bf e} of the main text). 
To derive the unitary transformation that will yield the simplified transport equivalent network shown in Fig.1{\bf f} of the main text, consisting of a single branch only, we note that one can apply successive transformations that each eliminate a single branch of the network. The total unitary transformation can then be written as a product of unitary transformations given by
\begin{equation}
{\hat U}_{tot} = {\hat U}_{1,2} \cdot {\hat U}_{2,3} \cdot \ldots \cdot {\hat U}_{P-1,P}
\label{eq:Utot}
\end{equation}
Here, ${\hat U}_{i,i+1}$ represents a unitary transformation that only affects sites in the $i$'th and $(i+1)$'th rows, and is given by
\begin{equation}
{\hat U}_{i,i+1} =
\begin{pmatrix}
{\hat 1} & 0 & 0 \\
0 & {\hat Q}_{i,i+1} & 0  \\
0 & 0 & {\hat 1}
\end{pmatrix}
\end{equation}
with ${\hat Q}_{i,i+1}$ being an $(2K \times 2K)$ matrix given by
\begin{equation}
{\hat Q}_{i,i+1} =
\begin{pmatrix}
{\hat D}_1(\alpha_i)  & 0 & 0 & \ldots \\
0 & {\hat D}_2(\alpha_i) & 0 & 0 & \ldots \\
0 & 0 & {\hat D}_3(\alpha_i) & 0 &  \ldots \\
0 & 0 & 0  & \ddots &  \ldots \\
\vdots & \vdots & \vdots  & \vdots &  {\hat D}_K(\alpha_i) \\
\end{pmatrix}
\end{equation}
where
\begin{equation}
{\hat D}_j(\alpha_i) =
\begin{pmatrix}
 \cos \alpha_i & -\sin \alpha_i  \\
\sin \alpha_i & \cos \alpha_i
\end{pmatrix}
\end{equation}
and $\alpha_i = \arctan \left[1/\sqrt{i} \right]$. Each block matrix ${\hat D}_j(\alpha_i)$ couples the $j'th$ sites in the $i'th$ and $(i+1)'th$ row.
In Fig.~\ref{fig:unitary_trans_1} we demonstrate the effect of the successive application of the matrices ${\hat U}_{i,i+1}$, to the Hamiltonian of the original network which is shown in Fig.~\ref{fig:unitary_trans_1}(a). The Hamiltonian $H^\prime = U_{1,2}^\dagger H U_{1,2}$ describes the network shown in Fig.~\ref{fig:unitary_trans_1}(b) (where we omitted the row of dots that are not connected to the leads after the transformation), while the Hamiltonian $H^{\prime \prime} = U_{2,3}^\dagger U_{1,2}^\dagger H U_{1,2} U_{2,3}$ describes the network shown in Fig.~\ref{fig:unitary_trans_1}(c). After $P-1$ transformations, we arrive at the network shown in Fig.~\ref{fig:unitary_trans_1}(d), which contains the fewest number of links and sites, and therefore can be considered the simplest transport equivalent network to the one shown in Fig.~\ref{fig:unitary_trans_1}(a). Note that in contrast to the unitary transformation considered in the context of Figs.1{\bf a}- {\bf c} of the main text, the above unitary transformation ${\hat U}_{tot}$ in Eq.(\ref{eq:Utot}) is not a simple direct sum of $SU(2)$ transformations.

\subsection{Networks with next-nearest-neighbor hoppings}

\begin{figure}[h]
 \begin{center}
\includegraphics[width=9.cm]{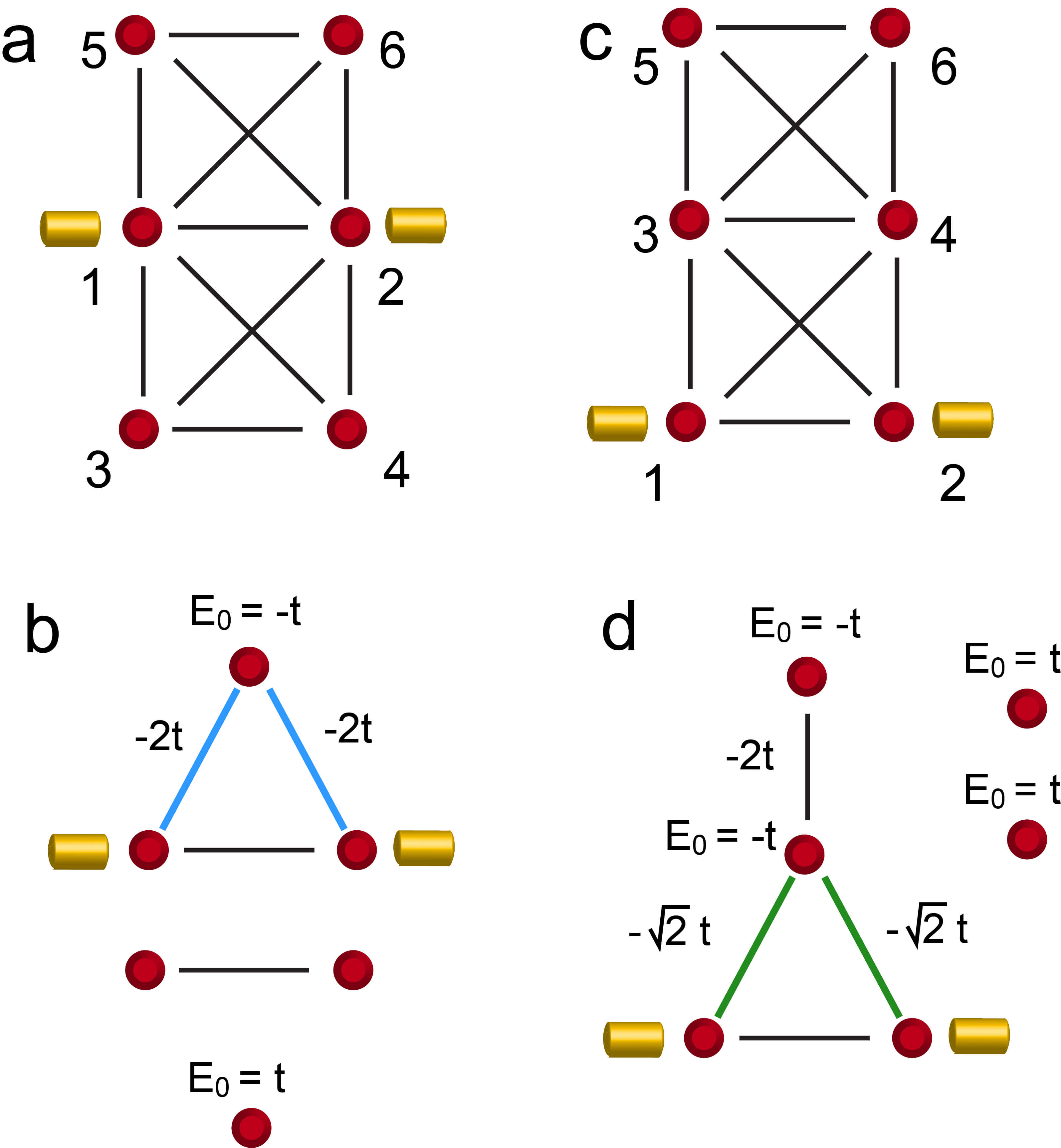}
\caption{Transport equivalent networks with next-nearest neighbor hoppings, and (a) leads attached to the center row, and (c) leads attached to the bottom row. Their transport equivalent networks are shown in (b) and (d), respectively. Here, blue (green) lines correspond to a hopping of $-2t$ ($-\sqrt{2}t$). Note that in the TENs, some sites now posses non-zero on-site energies $E_0$.}
\label{fig:criss_cross}
\end{center}
\end{figure}

Even for more complex networks containing next-nearest neighbor hoppings, as shown in Fig.~\ref{fig:criss_cross}, it is possible to find unitary transformations that lead to simplified transport equivalent networks.
For the case of leads attached to the center row, as shown in Fig.~\ref{fig:criss_cross}(a), the unitary transformation [using the numbering of sites shown in Fig.~\ref{fig:criss_cross}(a)] that relates the network in Fig.~\ref{fig:criss_cross}(a) to its transport equivalent network shown in Fig.~\ref{fig:criss_cross}(b) is given by  a product of unitary matrices
\begin{equation}
{\hat U} = {\hat U}_{1} \cdot {\hat U}_{2}
\end{equation}
where
\begin{equation}
{\hat U}_{1} =
\begin{pmatrix}
{\hat 1} & 0 & 0 \\
0 & {\hat D}(\alpha) & 0  \\
0 & 0 & {\hat D}(\beta)
\end{pmatrix}
\end{equation}
with ${\hat D}$ being a $(2 \times 2)$ matrix given by
\begin{equation}
{\hat D}(\alpha) =
\begin{pmatrix}
\cos \alpha & - \sin \alpha  \\
\sin \alpha  & \cos \alpha
\end{pmatrix}
\label{eq:Dalpha}
\end{equation}
and
\begin{equation}
{\hat U}_{2} =
\begin{pmatrix}
1 & 0 & 0 &0 &0 &0 \\
0 & 1 & 0 & 0 &0 &0 \\
0 & 0 & \cos \gamma & 0 & - \sin \gamma & 0 \\
0 &0 &0 & 1 & 0 & 0 \\
0 & 0 & \sin \gamma & 0 &  \cos \gamma & 0 \\
0 &0 &0 & 0 & 0 & 1 \\
\end{pmatrix}
\end{equation}
For $\alpha = \beta = \gamma = \pi/4$ the Hamiltonin matrix ${\hat H}^\prime  = {\hat U}^\dagger {\hat H} {\hat U}$ describes the system shown in Fig.~\ref{fig:criss_cross}(b). Interestingly enough, the transport equivalent network in Fig.~\ref{fig:criss_cross}(b) does not only possess renormalized hopping elements, but one of the sites also possesses a non-zero on-site energy $E_0$.
For the network with leads attached to the bottom row, Fig.~\ref{fig:criss_cross}(c), the unitary transformation [using the numbering of sites shown in Fig.~\ref{fig:criss_cross}(c)] leading to the transport equivalent network shown in Fig.~\ref{fig:criss_cross}(d) is given by
\begin{equation}
{\hat U}=
\begin{pmatrix}
{\hat 1} & 0 & 0 \\
0 & {\hat D}(\alpha) & 0  \\
0 & 0 & {\hat D}(\beta)
\end{pmatrix}
\end{equation}
with ${\hat D}(\alpha)$ given in Eq.(\ref{eq:Dalpha}), and $\alpha=\beta=\pi/4$.

\subsection{Networks with square or graphene lattice structure}

\begin{figure}[h]
 \begin{center}
\includegraphics[width=14.cm]{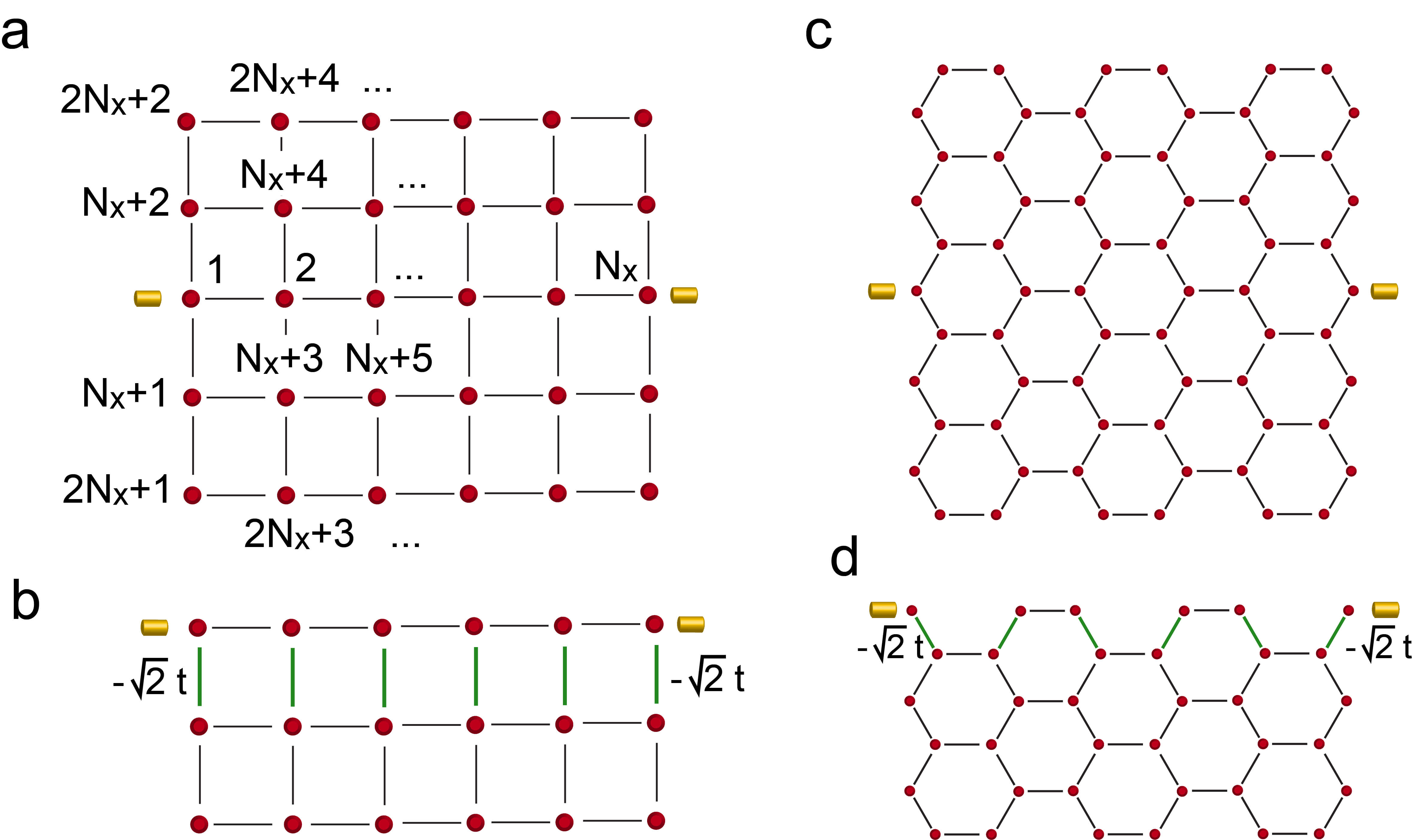}
\caption{(a) Network with a square lattice structure and nearest neighbor hopping only, and (b) its smallest TEN. (c) Network with a graphene lattice structure, and (d) its smallest TEN. Here, green lines correspond to a hopping of $- \sqrt{2}t$.}
\label{fig:square1}
\end{center}
\end{figure}

To find transport equivalent networks for networks with a square lattice symmetry and nearest neighbor hopping only, we consider the network shown in Fig.~\ref{fig:square1}(a). 
Using the numbering of sites shown in Fig.~\ref{fig:square1}(a), we find that the unitary transformation to obtain a TEN is given by
\begin{equation}
{\hat U} =
\begin{pmatrix}
{\hat 1} & 0 & 0 & \ldots \\
0 & {\hat D}(\alpha_1) & 0 & \ldots  \\
0 & 0 & {\hat D}(\alpha_2)& \ldots \\
0 & 0 & 0 & \ddots
\end{pmatrix}
\label{eq:square}
\end{equation}
with ${\hat D}$ being the same $(2 \times 2)$ matrix as given in Eq.(\ref{eq:Dalpha}), and ${\hat 1}$ being the $\left[ (N_x+2) \times (N_x+2) \right]$ identity matrix.  For $\alpha_i = \pi/4 \ \forall i$ we then obtain the TEN shown in Fig.~\ref{fig:square1}(b). Applying the same numbering to the network with a graphene lattice structure shown in Fig.~\ref{fig:square1}(c) allows us to use the same unitary transformation, ${\hat U}$ [Eq.(\ref{eq:square})] to arrive at the TEN shown in Fig.~\ref{fig:square1}(d).

\subsection{Networks with symmetric disorder}

\begin{figure}[h]
 \begin{center}
\includegraphics[width=12.cm]{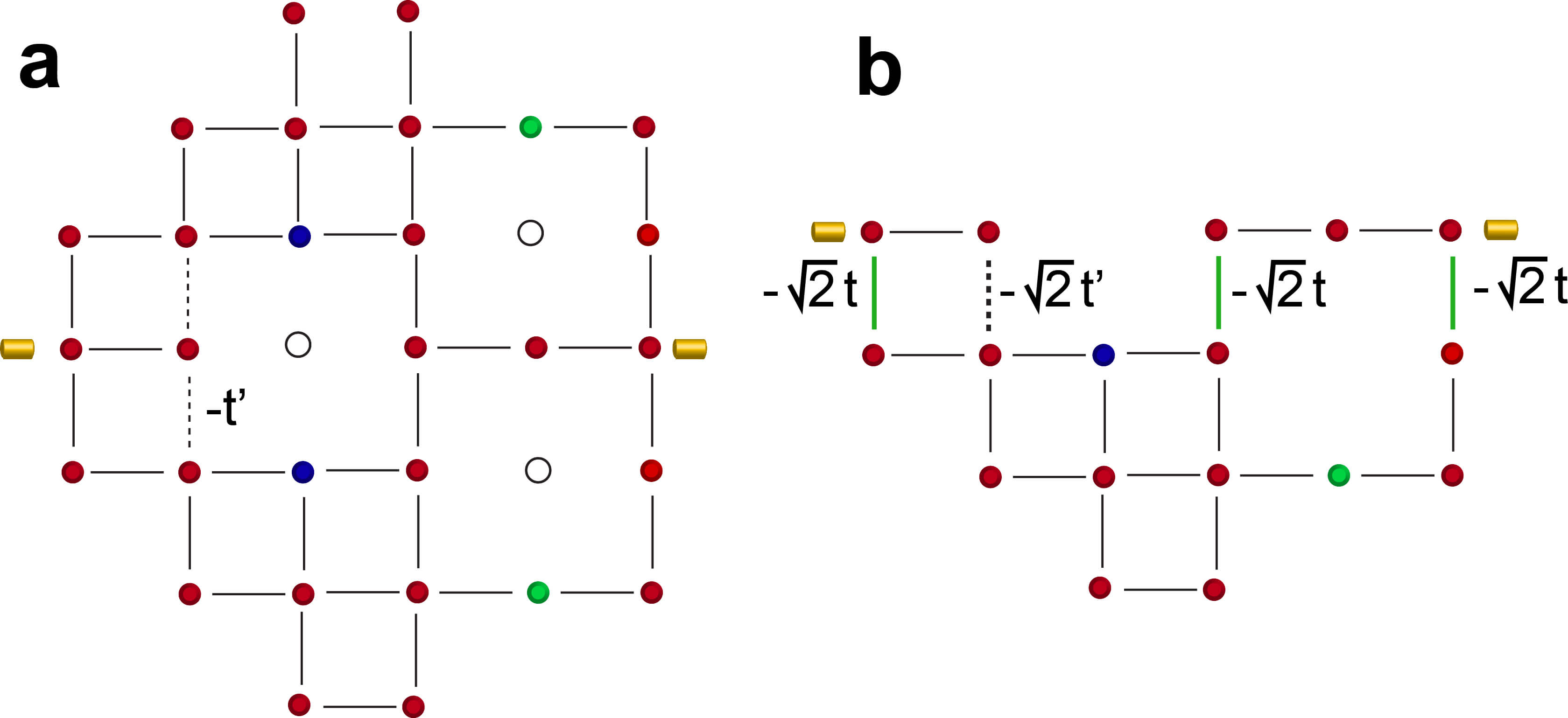}
\caption{(a) A network with disorder symmetric around the center row. Empty circles represent missing sites, green (blue) filled circles represents sites with the same on-site disorder, as represented by a potential scattering term and the dashed line represents disorder in the hopping element. (b) The smallest TEN of the network shown in (a). }
\label{fig:disorder}
\end{center}
\end{figure}

Disordered networks in general do not possess transport equivalent networks, since the disorder breaks the spatial symmetry of the network which is necessary for the derivation of transport equivalent networks. The exception to this rule, however, are disordered networks in which the disorder preserves the mirror symmetry of the network around the center row (we refer to such disorder as {\it symmetric disorder}). An example of such a network with symmetric disorder is shown in Fig.~\ref{fig:disorder}(a).
Here, empty circles represent missing sites, filled circles of the same color possess the same (disordered) on-site energy $E_0 \not = 0$, implying a local scattering potential $U_{i} = E_0$, and dashed lines represent disorder in the hopping elements. Using the same numbering of sites as in Fig.\ref{fig:square1}(a), and the same unitary transformation as in Eq.(\ref{eq:square}) with $\alpha_i=\pi/4 \ \forall i$, we obtain the TEN shown in Fig.~\ref{fig:disorder}(b).

\subsection{Three-Dimensional Networks}

\begin{figure}[h]
 \begin{center}
\includegraphics[width=12.cm]{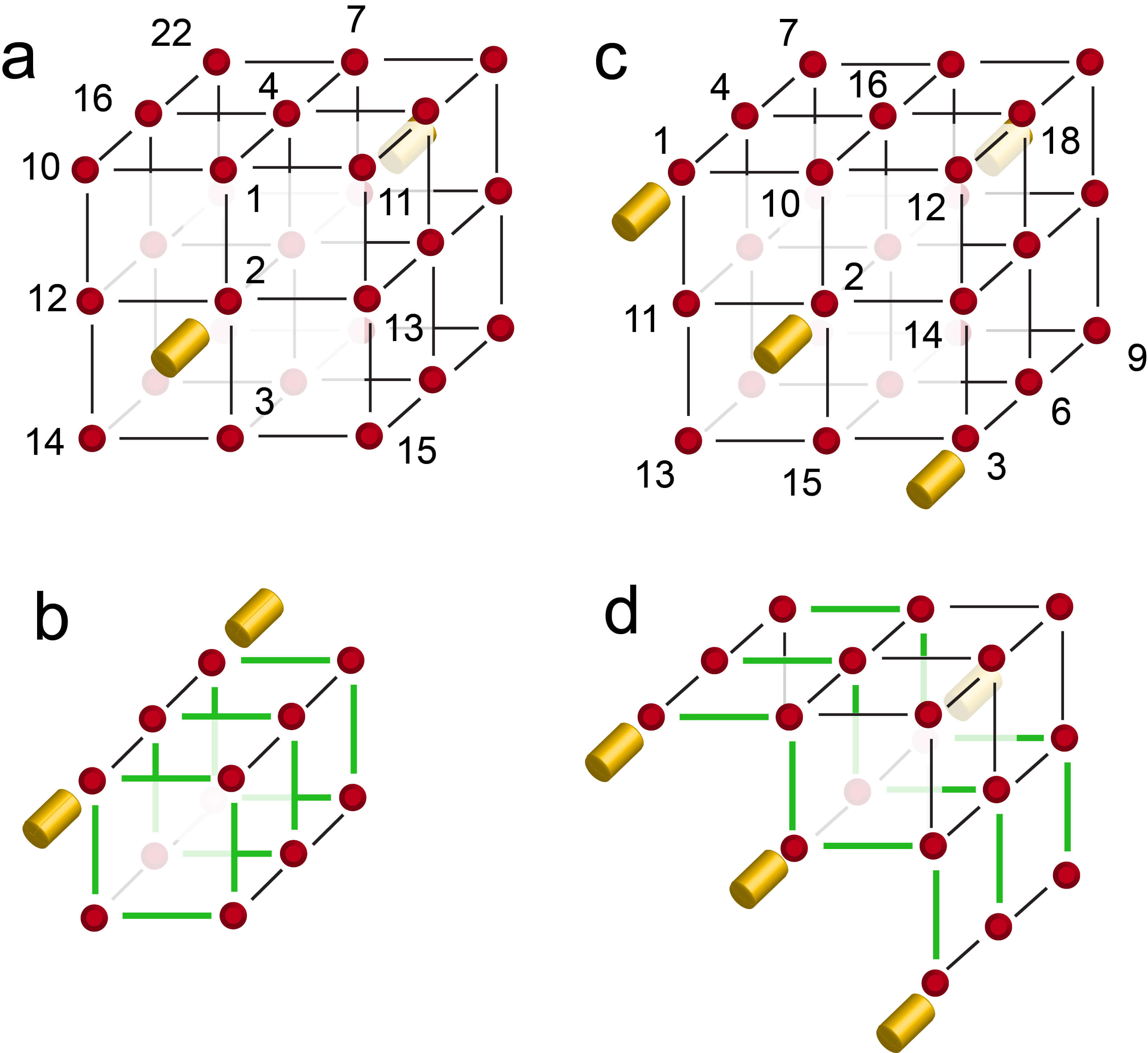}
\caption{(a) A three-dimensional network with single leads attached to its center sites, and (b) its smallest TEN. (c) A three-dimensional network with extended leads, and (d) its smallest TEN. Green lines correspond to a hopping of $-\sqrt{2}t$.}
\label{fig:3D}
\end{center}
\end{figure}
In order to find the smallest possible TEN to the three-dimensional network shown in Fig.~\ref{fig:3D}(a), we employ the unitary transformation [using the numbering of sites shown in Fig.~\ref{fig:3D}(a)]
\begin{equation}
{\hat U} = {\hat U}_1 \dot {\hat U}_2
\end{equation}
with
\begin{equation}
{\hat U}_1 =
\begin{pmatrix}
{\hat 1} & 0 & 0 & \ldots \\
0 & {\hat D}(\alpha) & 0 & \ldots  \\
0 & 0 & {\hat D}(\alpha)& \ldots \\
0 & 0 & 0 & \ddots
\end{pmatrix}
\label{eq:3D1}
\end{equation}
where ${\hat 1}$ is the $11 \times 11$ identity matrix describing the two lead sites and the center sites 1-9, and there are nine ${\hat D}(\alpha) $ matrices. Moreover,
\begin{equation}
{\hat U}_2 =
\begin{pmatrix}
{\hat 1} & 0 & 0 & 0 & 0 & 0 & 0 \\
0 & {\hat B}(\theta) & 0 & 0& 0 & 0 & 0 \\
0 & 0& {\hat B}(\theta)& 0& 0 &  0 &  0\\
0 & 0 & 0 & {\hat B}(\theta) & 0 & 0 & 0 \\
0 & 0& 0 & 0 & {\hat F}(\theta) &  0 & 0 \\
0& 0 &  0 & 0 & 0& {\hat F}(\theta)&  0\\
0 & 0 & 0 &0 & 0 & 0 & {\hat F}(\theta)  \\
\end{pmatrix}
\label{eq:3D2}
\end{equation}
where now ${\hat 1}$ is the $2 \times 2$ identity matrix describing the two lead sites, ${\hat B}(\theta)$ are $3 \times 3$ matrices given by
\begin{equation}
{\hat B}(\theta) =
\begin{pmatrix}
\cos \theta & 0 & -\sin \theta \\
0 & 1 & 0  \\
\sin \theta & 0 & \cos \theta \\
\end{pmatrix}
\label{eq:hatB}
\end{equation}
and ${\hat F}(\theta)$ are $6 \times 6$ matrices given by
\begin{equation}
{\hat F}(\theta) =
\begin{pmatrix}
1 & 0 & 0 & 0 & 0 & 0\\
0 & \cos \theta & 0 & 0 & 0 & -\sin \theta \\
0 & 0 & 1 & 0& 0 & 0 \\
0 & 0 & 0& 1& 0 & 0 \\
0 & 0 & 0 & 0& 1 & 0  \\
0 & \sin \theta & 0 & 0 & 0 & \cos \theta \\
\end{pmatrix}
\label{eq:hatB}
\end{equation}
The resulting TEN for $\alpha = \theta = \pi/4$ is shown in Fig.~\ref{fig:3D}(b).

For the network shown in Fig.~\ref{fig:3D}(c), we employ the unitary transformation [using the numbering of sites shown in in Fig.~\ref{fig:3D}(c)]
\begin{equation}
{\hat U} =
\begin{pmatrix}
{\hat 1} & 0 & 0 & \ldots \\
0 & {\hat D}(\alpha) & 0 & \ldots  \\
0 & 0 & {\hat D}(\alpha)& \ldots \\
0 & 0 & 0 & \ddots
\end{pmatrix}
\label{eq:3D1}
\end{equation}
where ${\hat 1}$ is the $11 \times 11$ identity matrix describing the two lead sites and the center sites 1-9, and there are nine ${\hat D}(\alpha) $ matrices. The resulting TEN for $\alpha = \pi/4$ is shown in Fig.~\ref{fig:3D}(d).

\section{Transport Equivalent Networks: {\it IV}-curves and current patterns}

\begin{figure}[h]
 \begin{center}
\includegraphics[width=15.cm]{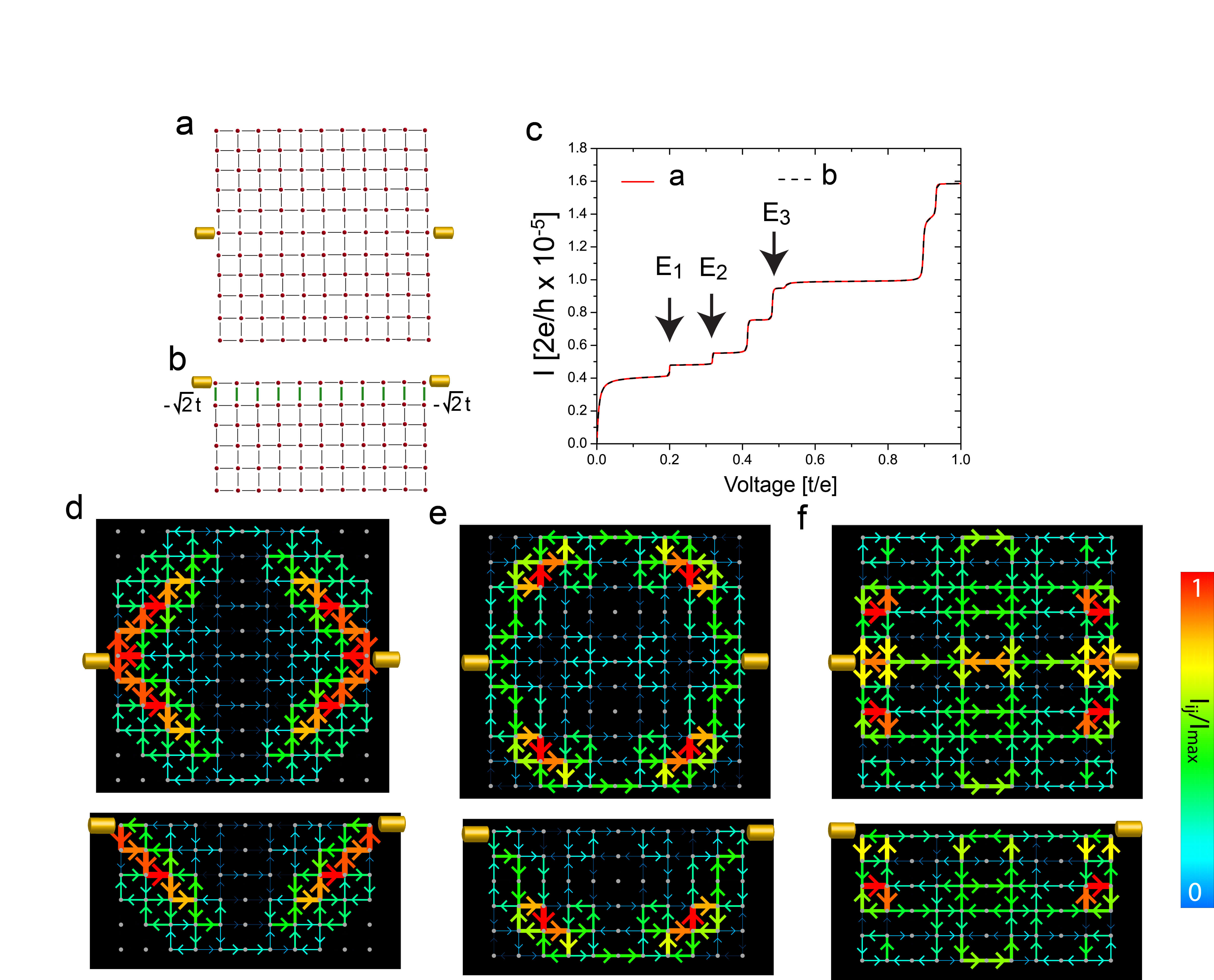}
\caption{(a) $(11 \times 11)$ network and (b) its transport equivalent $(11 \times 6)$ network shown in Fig.~3{\bf a} and 3{\bf b} of the main text, together with (c) their $IV$-curves. Spatial current patterns of the two networks carried by the states at energies (d) $E_1=0.1998t$, (e) $E_2=0.3179t$, and (f)$E_3=0.4824t$, as indicated by the arrows in (c) for $\Delta V = 0.01t /e$. These states are accessed by applying a gate voltage  $V_g=E_i/e$ to the network.}
\label{fig:spatial_patterns}
\end{center}
\end{figure}

\begin{figure}[h]
 \begin{center}
\includegraphics[width=7.cm]{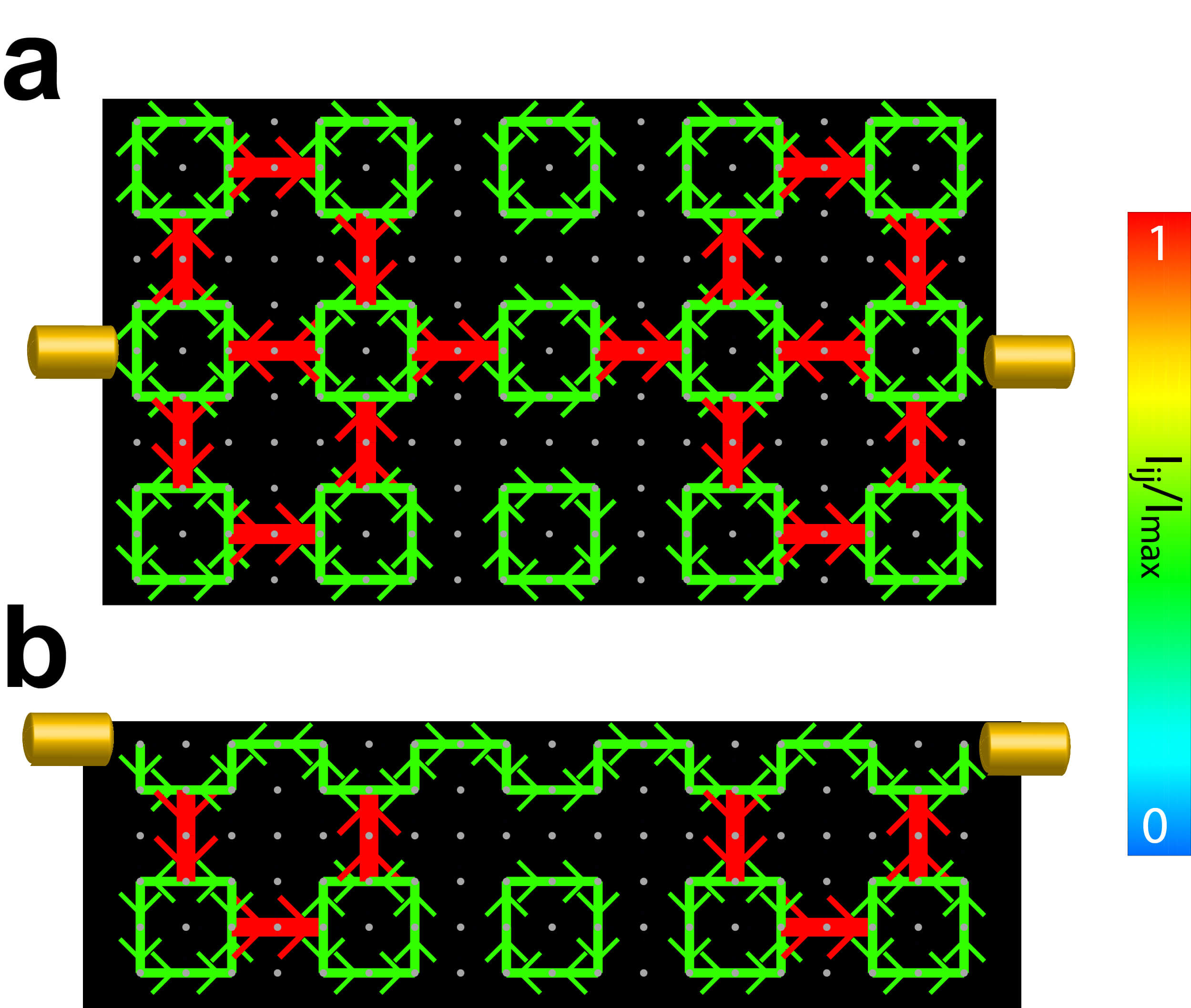}
\caption{Spatial pattern of the current carried by the $E=0$ state for $\Delta V = 0.01t /e$ in a (a) $(19 \times 11)$ network and (b) its transport equivalent $(19 \times 6)$ network.}
\label{fig:19x11}
\end{center}
\end{figure}

We showed in Fig.~3{\bf c} and 3{\bf d} of the main text that the spatial patterns of the current carried by the $E=0$ state in two transport equivalent networks [see Fig.~\ref{fig:spatial_patterns}(a) and (b)] show a high degree of similarity. This similarity also holds for the spatial pattern of currents carried by other energy states, as shown in Figs.~\ref{fig:spatial_patterns}(d) - (f).
The corresponding energy states appear as jumps in the networks' $IV$ curves, as indicated by arrows in Figs.~\ref{fig:spatial_patterns}(c). Moreover, the similarity of the current patterns also holds for square-lattice networks with $N_x \not = N_y$, as shown in Fig.~\ref{fig:19x11}.
Since $N_x \not = N_y$, the spatial pattern of the current carried by the $E=0$ state in the $(19 \times 11)$ networks [Fig.~\ref{fig:19x11}(a)] exhibits a more complex form. Nevertheless, the spatial current pattern in the $(19 \times 6)$ TEN [Fig.~\ref{fig:19x11}(b)] again exhibits a close similarity.

\section{Spatial Current Patterns in transport equivalent networks in the presence of an electron-phonon interaction}

\begin{figure}[h]
 \begin{center}
\includegraphics[width=15.cm]{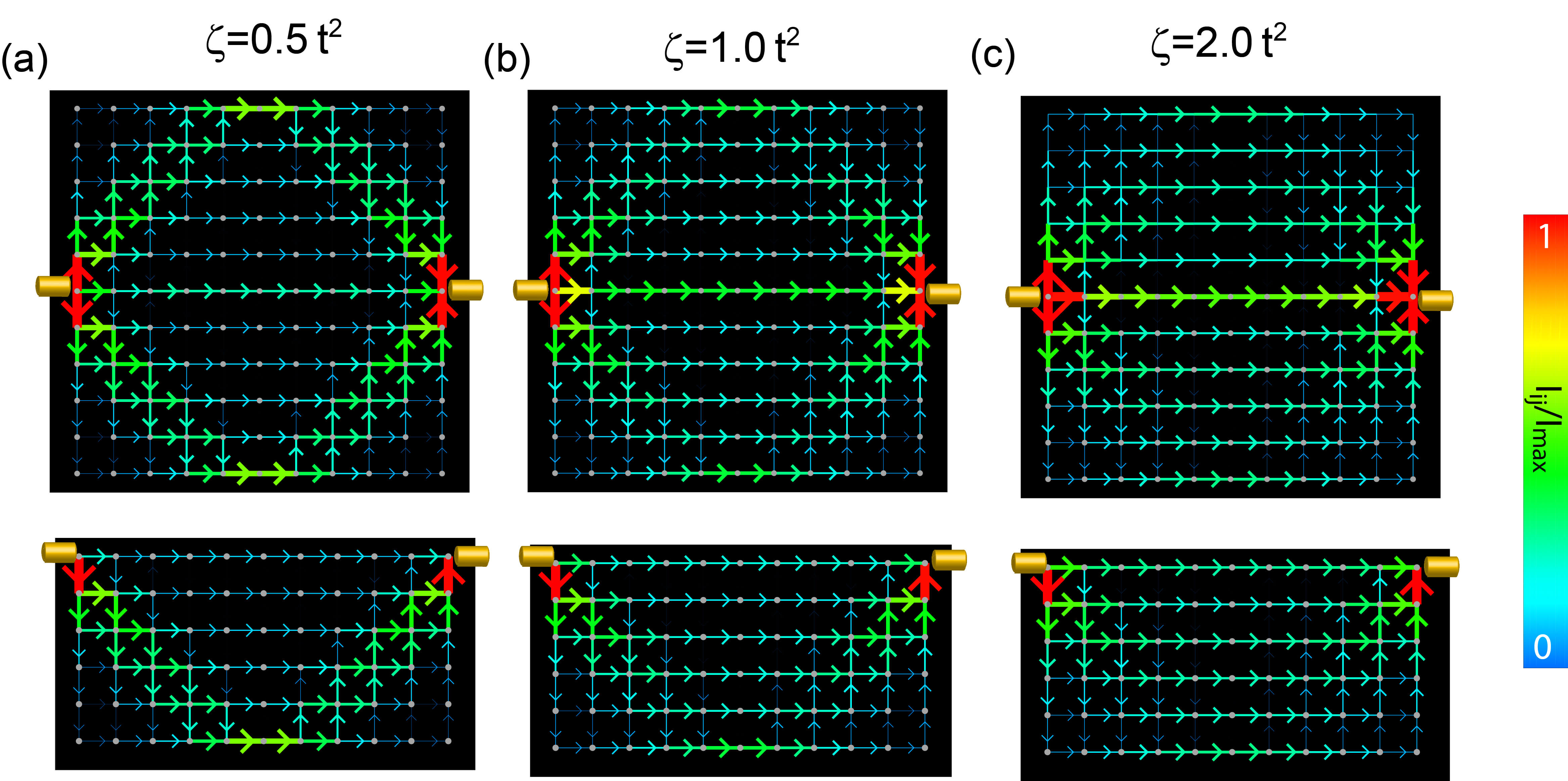}
\caption{Evolution of the spatial current patterns of the current carried by the $E=0$ state for $\Delta V = 0.01t /e$ with increasing $\zeta$ for the two TENs shown in Figs.3{\bf a} and 3{\bf b} of the main text: (a) $\zeta=0.5 t^2$, (b) $\zeta=1.0t^2$, and (c) $\zeta=2.0t^2$.}
\label{fig:spatial_patterns_zeta}
\end{center}
\end{figure}
To consider the effects of dephasing on the current pattern in the square-lattice network and its TEN shown in in Figs.~\ref{fig:square1}(a) and (b) (see also Figs.3{\bf a} and {\bf b} of the main text) we take the electron-phonon interaction to be of the form shown in Eq.(6) of the main text, where in the original network symmetric sites are coupled to the same two phonons modes. Specifically, using the numbering of sites shown in Fig.~\ref{fig:square1}(a), the pairs of symmetric sites are given by $i=pN_x+(2k-1)$ and $j=pN_x+2k$ ($p,k=1,2,...)$. Using the unitary transformation of Eq.(\ref{eq:square}), the electron-phonon in the TEN is local, as described by the Hamiltonian $H^\prime_{ph}$ in Eq.(7) of the main text. We showed in Figs.3 and 4 of the main text that the spatial current patterns of TENs exhibit a close similarity both in the quantum limit $\zeta =0$ (Fig.3) and in the classical limit $\zeta \rightarrow \infty$ (Fig.4). However, this close similarity also holds for intermediate values of $\zeta$, as shown in Fig.~\ref{fig:spatial_patterns_zeta}. This implies that the global transport equivalence of networks is reflected in a close similarity of the local transport properties -- as reflected in the spatial current patterns -- over the entire range from quantum to classical transport.

\section{Equivalent Transport in Classical Networks}

\begin{figure}[h]
 \begin{center}
\includegraphics[width=15.cm]{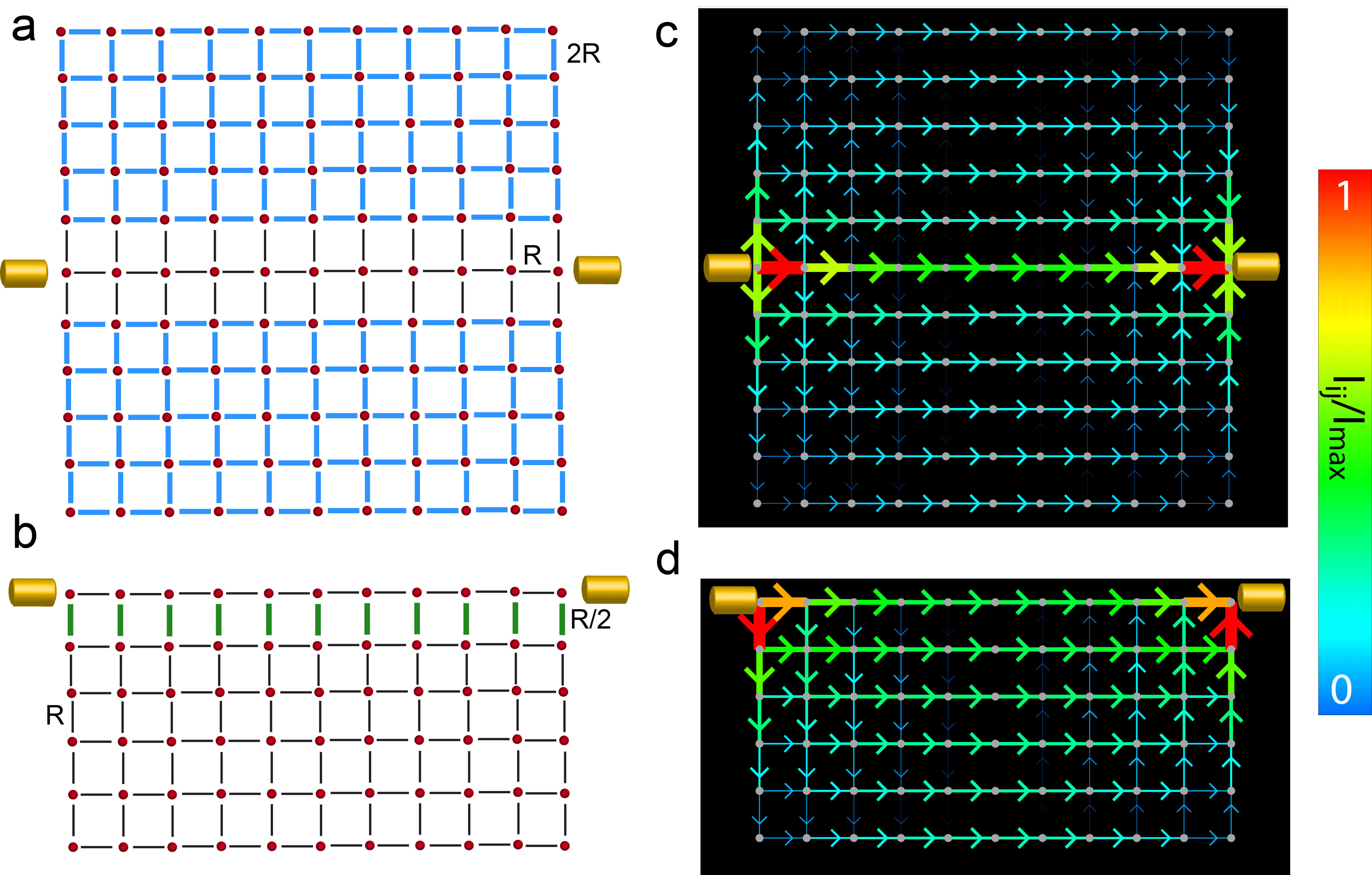}
\caption{(a),(b) Transport equivalent classical resistor networks that are obtained in the limit $\zeta \rightarrow \infty$ from the networks shown in Figs.3 of the main text. The thick blue lines represent resistors with resistance $2R$, the thin black lines a resistor with resistance $R$, and the thick green lines represent resistors with resistance $R/2$. (c),(d) Spatial flow of currents through the networks in (a),(b), respectively. }
\label{fig:class_networks}
\end{center}
\end{figure}

We showed in the main text that transport equivalent networks remain transport equivalent over the entire range from the quantum, $\zeta \rightarrow 0$ to the classical $\zeta \rightarrow \infty$ limit. Here, we show that in the classical limit, $\zeta \rightarrow \infty$, the transport properties of the networks map onto those of classical resistor networks, which therefore are also transport equivalent. Specifically, in Figs.~\ref{fig:class_networks}(a) and (b) we present the two classical resistor networks that possess identical transport properties to the networks shown in Figs.3{\bf a} and 3{\bf b} of the main text in the $\zeta \rightarrow \infty$ limit.
This equivalence of transport properties between the two networks in the limit of $\zeta \rightarrow \infty$, and the two classical resistor networks follows from a comparison of the spatial current patterns. In particular, the spatial current pattern  of the two classical resistor networks \cite{Wu04}, shown in  Figs.~\ref{fig:class_networks}(c) and (d), are identical to those of the two transport equivalent networks in the limit $\zeta \rightarrow \infty$ shown in Figs.~4{\bf c} and 4{\bf d} of the main text. Moreover, one can easily check that the classical network shown in ~\ref{fig:class_networks}(b) has an equivalent resistance to the network shown in ~\ref{fig:class_networks}(a). This demonstrates that the concept of an equivalent resistance, or of transport equivalent networks, can be extended to the entire range from the quantum to the classical transport limit.